\documentstyle[prd,aps]{revtex}
\newcommand{\be}{\begin{equation}}
\newcommand{\ee}{\end{equation}}
\newcommand{\bn}{\begin{eqnarray}}
\newcommand{\en}{\end{eqnarray}}
\newcommand{\bd}{\begin{displaymath}}
\newcommand{\ed}{\end{displaymath}}
\newcommand{\bnn}{\begin{eqnarray*}}
\newcommand{\enn}{\end{eqnarray*}}
\begin{document}
\title{Ground State Energy of Massive Scalar Field in the Global
Monopole Background}
\author{ E.R. Bezerra de Mello \thanks{e-mail:
emello@fisica.ufpb.br}, V.B. Bezerra \thanks{e-mail:
valdir@fisica.ufpb.br}, and N.R. Khusnutdinov \thanks{On leave from
Kazan State Pedagogical University, Kazan, Russia; e-mail:
nail@dtp.ksu.ras.ru}}
\address{ Departamento de F\'{\i}sica, Universidade Federal da
Para\'{\i}ba, \\ 
Caixa Postal 5008, CEP 58051-970 Jo\~ao Pessoa, Pb, Brazil}
\date{\today}
\maketitle
\begin{abstract}
We consider the ground state energy of scalar massive field in
the spacetime of a pointlike global monopole. Using
zeta function regularization method we obtain the heat kernel
coefficients for this system. We show that the coefficient $B_1$
contains additional contribution due to the non-trivial topological
structure of the spacetime. Taking into account the heat kernel
coefficients we obtain that the ground state energy of the scalar
field is zero. We also discuss our result using dimensional 
considerations. 
\end{abstract}
\section{Introduction}
Different types of topological objects may have been formed
during Universe expanding, such as domain walls, cosmic
strings and monopoles \cite{Kibble}.  These topological deffects
appeared due to breakdown of local or global gauge symmetries.
Global monopoles are created due to phase transition when a
global gauge symmetry is broken and they may have been important
for cosmology and astrophysics. The process of global monopole
creation is accompanied by particles production \cite{Lousto}.
Grand Unified Theory predicts great number of these objects in
the Universe \cite{Preskill} but the problem may be avoided
using inflationary models. From astrophysical point of view
there is at most one global monopole in the local group of
galaxies \cite{HiscockPRL}.  

The spacetime of a global monopole in a $O(3)$ broken symmetry model
has been investigated by Barriola and Vilenkin
\cite{BarriolaVilenkin}. They have shown that far from the
compact monopole's core the spacetime is approximately described
by spherical symmetry metric and with additional solid angle
deficit (see also Ref. \cite{HarariLousto}). There is a global
monopole with regular core and another one with 
pointlike singular core which we study in this paper and also
call regular monopole. The spacetime is not locally flat, even
for the case of pointlike global monopole spacetime.   

The analysis of quantum fields on the global monopole background
have been considered in
Refs. \cite{FieldsOnGlobalMonopole,Mazzitelli}. It was shown, taking
into account only dimensional and conformal consideration
\cite{FieldsOnGlobalMonopole}, that the vacuum expectation value of
the energy momentum tensor of conformal massless scalar, spinor and
vector fields on this background has the following general structure 
\bd
T^i_k = S^i_k{\hbar c\over r^4}\ ,
\ed
where the quantities $S^i_k$ depend on solid angle deficit and spin
of the fields. For scalar field this tensor was investigated in
great details in the paper by Mazzitelli and Lousto
\cite{Mazzitelli}.  

The energy-momentum tensor has nonintegrable singularity at
origin and therefore the ground state energy cannot be found by
integrating the energy density. The same problem also appears for
cosmic string spacetime \cite{FieldsOnCosmicString} and in
Minkowsky spacetime with boundary condition on dihedral angle
\cite{Dihedrial}. For the cosmic string spacetime this problem 
was considered in Refs. \cite{Dima,KhusnutdinovBordag} by 
using another approaches for calculation of ground state energy.
In spite of the energy density has singular form it has been
found that the ground state energy of massive scalar field is
zero. For infinitely thin cosmic string this may be explained
taking into account only dimensional considerations (see below
Sec.\ref{Last}).  

Nontrivial topological structure of spacetime reveals itself via
some contribution in the heat kernel coefficients. In the case of
a cosmic string spacetime it appeared in the heat kernel
coefficient $B_1$ \cite{Kac} as additional contribution
along with usual volume and boundary terms
\cite{Dima,Klaus,KhusnutdinovBordag}. But this question has not
been yet investigated for global monopole spacetime.  

Nontrivial topological structure of spacetime leads to a number of
interesting effects which are forbidden in flat space. For
example, there appear additional self-interacting force
which is non-zero even for particle in the rest. This force has
been investigated in Refs. \cite{Linet,Bezerra} for cosmic string
and global monopole spacetimes respectively.

In this paper we would like to discuss the ground state
energy of scalar massive field with arbitrary nonconformal coupling
in the background of pointlike global monopole spacetime and
calculate topological contribution to the heat kernel
coefficients.  

In the framework of zeta function regularization method 
\cite{ZetaFunction} (see also \cite{ElizaldeBook}) the
ground state energy of scalar massive field is given by 
\be
E(s) = {1\over 2}M^{2s}\zeta_A(s-{1\over 2})\ , 
\label{GroundEnergy}
\ee 
which is expressed in terms of the zeta function $\zeta_A $ of the
Laplace operator $\hat A = -\triangle + \xi {\cal R} +m^2$ on three 
dimensional spatial section of spacetime. Here, parameter $M$ with
dimension of mass keeps right dimension of energy. For calculation
and renormalization of the ground state energy we use the approach
which was suggested and developed in Refs. \cite{Method,Method2}. 

Zeta function of Laplace operator on the global monopole background
has been considered in detail by Bordag, Kirsten and Dowker
in Ref. \cite{BordagKirstenDowker} using the method given in
Refs. \cite{Method,Method2}. There the general mathematical
structure of zeta function and the heat kernel coefficients on
the generalized cone were obtained. We shall rederive some
formulas in our concrete case because the main emphasis of the
present paper is ground state energy which was not considered in
Ref. \cite{BordagKirstenDowker}. It is necessary also in order to
compare our considerations with the general formulas of a nonsingular 
background and separate the additional topological contributions. 

The organization of the paper is as follows. In Sec.\ref{Geometry} we
give some geometrical data about global 
monopole spacetime which will be needed. In Sec.\ref{Zeta}, the zeta
function of the Laplace operator on three dimensional section of
a global monopole spacetime is calculated and the heat kernel
coefficients are obtained including topological contribution. In
Sec.\ref{GrEn}, the ground state energy of massive scalar field with
arbitrary nonconformal coupling on global monopole background is
considered. In last Sec.\ref{Last}, we discusse our results.
The signature of the spacetime, the sign of Riemann and Ricci
tensors are the same as in Christensen paper \cite{Christensen}.
We use units $\hbar = c= G = 1$. 
\section{The Geometry}\label{Geometry}
Global monopoles are heavy objects probably formed in the early
Universe by the phase transition which occur in a system
composed by a self - coupling scalar field triplet $\phi^a$
whose original global symmetry $O(3)$ is spontaneously
broken to $U(1)$. 

The simplest model which gives rise a global monopole is
described by the Lagrangian density below
\bd
L= {1\over 2}(\partial_l \phi^a) (\partial^l \phi^a) - {\lambda
\over 4} (\phi^a \phi^a - \eta^2)^2\ .
\ed
Coupling this matter field with the Einstein equation, Barriola
and Vilenkin \cite{BarriolaVilenkin} have shown that the effect
produced by this object in the geometry can be approximately
represented by a solid angle deficit in the $(3+1)$ -
dimensional spacetime, whose line element is given by
\be
ds^2=-dt^2 +\alpha^{-2}dr^2 + r^2(d\theta^2 + \sin^2\theta
d\varphi^2)\ ,\label{Metrica}
\ee
where the parameter $\alpha^2 = 1-8\pi\eta^2$ is smaller than
unity and is connected with the symmetry breaking energy scale
$\eta$. The solid angle has the value $4\pi\alpha^2<4\pi$. Note that
the spacetime given by (\ref{Metrica}) is not flat. Nonzero
components of Riemann and Ricci tensors, and scalar curvature read 
\bd
{\cal R}^{\theta\varphi}_{.\ .\theta\varphi}={\cal
R}^{\theta}_{\theta}= {\cal R}^{\varphi}_{\varphi}={1-\alpha^2
\over r^2}\ ,\ {\cal R}={2(1-\alpha^2)\over r^2}\ .
\ed
For further application consider extrinsic curvature tensor on
the sphere of radius $R$ around the origin
\bd
K_{ij}=\nabla_i N_j\ .
\ed
Here $N_j$ is outward unit normal vector with coordinates
$N_j=(0,\alpha,0,0)$. This tensor has two nonzero components
\bd
K^{\theta}_{\theta} = K^{\varphi}_{\varphi}={\alpha \over R}\ . 
\ed
\section{Zeta Function and Heat Kernel Coefficients}\label{Zeta}
In order to calculate the ground state energy given by 
(\ref{GroundEnergy}) we have to obtain the zeta function of the
operator $\hat A$ in the neighbourhood of point $-1/2$. For
calculation of zeta function we follow to Ref. \cite{Method2}
The zeta function of the operator $\hat A = -\triangle + \xi {\cal R}
+ m^2$ is defined in terms of the sum over all eigenvalues of this
operator by 
\bd
\zeta_A(s-{1\over 2}) = \sum_{(n)}(\lambda_{(n)}^2 +
m^2)^{1/2-s}\ . 
\ed
Here $\lambda_{(n)}^2$ is the eigenvalue of operator 
$\hat B = \hat A - m^2$. The eigenfunctions which are regular at
the origin have the form 
\bd
\Phi ({\bf r}) = \sqrt{\lambda \over \alpha r} Y_{lm}(\theta,
\varphi) J_\mu ({\lambda \over \alpha}r)\ , 
\ed
where $Y_{lm}$ is the spherical harmonics and $J_\mu$ the Bessel
function of the first kind with index 
\be
\mu = {1\over \alpha}\sqrt{(l+{1\over 2})^2 + 2(1 - \alpha^2)(\xi -
{1\over 8})}\ . 
\label{index}
\ee
The eigenvalues $\lambda_{l,j}$ can be found by some boundary
condition imposed on this function. Let us consider the
Dirichlet boundary condition at the surface of a sphere of
radius $R$ 
\be
\sqrt{\lambda_{l,j}} J_\mu({\lambda_{l,j} \over \alpha}R)=0\ .
\label{Boundary} 
\ee 
So, the zeta function reads now as:
\bd
\zeta_A(s-{1\over 2}) = \sum_{l=0}^{\infty} \sum_{j=0}^{\infty}
(2l+1)(\lambda_{l,j}+m^2)^{1/2 -s}\ .
\ed
The solutions $\lambda_{l,j}$ of equation (\ref{Boundary}) can
not be found in closed form. For this reason we use the method
suggested in Refs. \cite{Method,Method2} which allows us to
express the zeta function in terms of the eigenfunctions. According
to this approach the sum over $j$ may be converted into contour
integral in complex $\lambda$ plane using the princip of
argument, namely 
\bd
\zeta_A(s-{1\over 2}) = \sum_{l=0}^\infty (2l + 1) \int_\gamma
d\lambda^2 (\lambda^2 + m^2)^{1/2 - s}{\partial \over \partial
\lambda} \ln \lambda^{-\mu} J_\mu ({\lambda \over
\alpha} R)\ ,
\ed
where the contour $\gamma$ runs counterclockwise and must
enclose all solutions of eq. (\ref{Boundary}) on positive real
axis. Next one shifts the contour to the imaginary axis. Doing
this we obtain the following formula for the zeta function (see
\cite{Method2} for details) 
\be
\zeta_A(s-{1\over 2}) = -{\cos \pi s\over \pi}
\sum_{l=0}^\infty (2l + 1) \int_m^\infty dk^2 (k^2 - m^2)^{1/2 -
s}{\partial \over \partial k} \ln k^{-\mu} I_\mu ({k \over
\alpha} R)\ .
\label{Main} 
\ee
Here $I_\mu$ is the modified Bessel function obtained from
$J_\mu$ in imaginary axis $(\lambda = ik)$. Next we use the
uniform expansion for the Bessel function $I_\mu(\mu z)$ as
below 
\be
I_{\mu}(\mu z) = \sqrt{\frac{t}{2\pi \mu}} e^{\mu \eta (z)}
\left\{ 1 + \sum_{k=1}^\infty \frac{u_k(t)}{(\mu)^k} \right\}\ ,  
\label{UniformExpansion}
\ee
where $t = 1/\sqrt{1 + z^2}\ ,\ \eta (z) = \sqrt{1 + z^2} + \ln
(z/(1 + \sqrt{1 + z^2}))$ and $z = kR/\mu\alpha$. The firsts few 
coefficients $u_k(t)$ and recursion relations for higher ones are
listed in \cite{Abramowitz}. This uniform expansion leads to power 
series over $m$, and term $u_N$ gives the contribution $\sim
1/m^{3-N}$. We shall make the calculations up to $N=3$. Using the
formula 
\bd
\int_1^\infty dx (x^2 - 1)^{1/2 -s} x (1 + x^2/\gamma^2 )^{-p/2}
= \frac{\Gamma (\frac{3}{2} - s) \Gamma (s + \frac{p - 3}{2})}{
2\Gamma (\frac{p}{2})} \gamma^p (1 + \gamma^2 )^{-s -
\frac{p-3}{2}}\ , 
\ed
we obtain the following expansion for the zeta function
\bn
\zeta_A(s-{1\over 2})&=& {m^{-2s} \over
(4\pi)^{3/2}}{4\pi^{3/2}m\beta /\alpha \over \Gamma (s - 1/2)}
\left\{ \sum_{l=0}^\infty (2l+1) \left[ 
\frac{\Gamma (s - 1)}{\sqrt{\pi}} {}_2F_1
-\frac{\alpha\mu}{\beta }\Gamma (s -\frac{1}{2}) \right]\right.
\nonumber \\ 
&-& {\alpha \over 2\beta} Z(0,s - \frac{1}{2}) -
\frac{\alpha^2}{4\beta^2 \sqrt{\pi}} \left[ Z(0,s) -
\frac{10}{3} Z(2, s + 1) \right] \nonumber \\ 
&-& \frac{\alpha^3}{8\beta^3} \left[ Z(0, s+ \frac{1}{2}) -
6Z(2, s+ \frac{3}{2}) + \frac{5}{2} Z(4, s+ \frac{5}{2})\right]
\label{ZetaN} \\ 
&-& \frac{\alpha^4}{96\beta^4 \sqrt{\pi}} \left[ 
25 Z(0, s + 1)\phantom{\frac{1}{2}} - \frac{1062}{5}Z(2, s + 2)
+ \frac{884}{5} Z(4, s + 3) \right. \nonumber \\
&-& \left.\left.\frac{1768}{63} Z(6, s + 4)
\right] + \dots \right\}\ . \nonumber
\en
Here ${}_2F_1 = {}_2F_1 (-{1\over 2}, s -1; {1\over 2}; -\left({\mu
\alpha \over \beta})\right)^2$ is the hypergeometric function;
$\beta = mR$ and 
\be
Z(p,q) = \Gamma (q) \sum_{l=0}^\infty {2l+1\over (1 + \alpha^2 
\mu^2/\beta^2)^q}\left(\frac{\alpha \mu}{\beta}\right)^p\ .
\label{Zet}
\ee  
The first expression in (\ref{ZetaN}) given in terms of
hypergeometric function 
\be
T(s)=\sum_{l=0}^\infty (2l+1) \left[ \frac{\Gamma (s -
1)}{\sqrt{\pi}} {}_2F_1 -\frac{\alpha\mu}{\beta }\Gamma (s
-\frac{1}{2}) \right]  \ ,
\label{F21}
\ee
can be expressed in terms of the same function given in Eq.
(\ref{Zet}). Indeed, one can use analytical continuation of the
hypergeometrical function \cite{Abramowitz}
\bnn
&& {}_2F_1\left(-\frac{1}{2}, s-1; \frac{1}{2};
-\left(\frac{\alpha\mu}{\beta}\right)^2 \right) 
     = 
\frac{\alpha\mu}{\beta} \frac{\Gamma (1/2) \Gamma (s -
1/2)}{\Gamma (s-1)} \\ 
&+&
\frac{\Gamma (1/2) \Gamma (1/2 - s)}{\Gamma (-1/2) \Gamma (3/2 -
s)} \left( 1 + \left(\frac{\alpha\mu}{\beta} \right)^2
\right)^{1-s} {}_2F_1\left(1, s-1; s + \frac{1}{2}; \frac{1}{1 +
\left(\frac{\alpha\mu}{\beta}\right)^2 }\right) \ . 
\enn
So, the first term in the rhs. of the above equation cancels the
second one, divergent term in the sum (\ref{F21}) which is
due to term $k^{-\mu}$ in (\ref{Main}) (see Ref. 
\cite{KhusnutdinovBordag}).  Next, one can use power series
expansion for the hypergeometric function because its argument
$1/(1 + (\alpha\mu/\beta)^2)$ is always smaller then unity, so
we arrive at 
\be
T(s)={1\over 2 \sqrt{\pi}}\Gamma (s-1/2) \sum_{l=0}^\infty
{Z(0,n+s-1) \over \Gamma (n+s +1/2)}\ . 
\label{T(s)}
\ee
Therefore for calculation of the zeta function we have to obtain 
an analytical continuation to the series $Z(p,q)$. In fact we may
consider only 
\be
Z(0,q) = \Gamma (q)\sum_{l=0}^\infty {2l+1\over (1 +
\alpha^2\mu^2/\beta^2)^q}\ ,
\label{Zet0}
\ee
because the other functions with $p=2,4,6,..$ can be expressed in
terms of $Z(0,q)$ only. Substituting the value for $\mu$ given in
(\ref{index}) into (\ref{Zet0}) we obtain 
\bd
Z(0,q)=2\Gamma (q)\beta^{2q}\sum_{l=0}^\infty {l+1/2\over
((l+1/2)^2 + b^2)^q}\ ,
\ed
where $b^2 = \beta^2 + 2(1-\alpha^2)(\xi -1/8)$. This series
is convergent for $\Re q >1$. For analytical continuation of this
function to the domain $\Re q\leq 1$ let us consider the series
below 
\bd
F(q,a,b^2) = \sum_{l=0}^\infty {1 \over ((l+1/2)^2 + b^2)^q}\ . 
\ed
This series has been considered in great details by Elizalde
\cite{Elizalde}. He found analytical continuation of this series
which for great $b$ reads 
\bnn
F(q,a,b^2) &\simeq& {b^{-2q}\over \Gamma (q)} \sum_{l=0}^\infty {(-1)^l
\Gamma (l+q)\over l!} b^{-2l} \zeta_H(-2l,a) + {\sqrt{\pi}
\Gamma (q-1/2) \over 2\Gamma (q)}b^{1 -2q} \nonumber \\
&-& {2\pi b^{-1/2 - q}\over \Gamma (q)} \sum_{n=1}^\infty
n^{q-1/2}\cos (2\pi n a) K_{q-1/2} (2\pi n b)\ .
\enn
Here $\zeta_H$ is the Hurwitz zeta function and $K_n$ is the
modified Bessel function. Differentiating this series with
respect to $a$ and putting $a=1/2$ we obtain the analytical
continuation we need which is the following
\bd
\sum_{l=0}^\infty {l+1/2 \over ((l+1/2)^2 + b^2)^q} \simeq {b^{2-2q}
\over 2(q-1)} + \sum_{l=0}^\infty {(-1)^l \Gamma (l+q) \over l!
\Gamma (q)} b^{-2q - 2l}\zeta_H(-1 -2l,1/2)\ .
\ed
Taking into account this expression we obtain analytical
continuation for function $Z(0,q)$:
\be
Z(0,q)\simeq \left({b^2\over \beta^2}\right)^{-q} \left\{ b^2 \Gamma
(q-1) + 2 \sum_{l=0}^\infty {(-1)^l \over l!} \Gamma (l+q)
b^{-2l} \zeta_H(-1-2l,1/2) \right\}\ ,
\label{Zet0An}
\ee
where $b^2/\beta^2 = 1 + 2(1-\alpha^2)(\xi - 1/8)/\beta^2$. This
function has simple poles in integer numbers $q = 1,0,-1,-2,...$
.  In order to calculate zeta function up to degree $m^0$ we
need only two terms from series (\ref{Zet0An}) in which 
$\zeta_H(-1,1/2) =1/24\ , \zeta_H(-3,1/2) = - 7/960$ and three
terms of $T(s)$  which is given by Eq.(\ref{T(s)}).  

Putting this expression into (\ref{T(s)}), and (\ref{ZetaN}) and
expanding over $1/\beta = 1/mR \ll 1$ and $s$, and collecting
terms with similar degree on the mass $m$ up to $m^0$ (we cannot
here collect higher orders of $m$ because we used 
uniform expansion up to this power) we get 
\bnn
\zeta_A (s-1/2) &=& {m^{-2s} \over (4\pi )^{3/2}} \left\{
\left[{4\pi R^3 \over 3\alpha}\right]m^4 {\Gamma (s-2)\over
\Gamma (s-1/2)} + \left[-2\pi^{3/2}R^2\right]m^3{\Gamma (s-3/2)
\over \Gamma (s-1/2)} \right.\\
&+& \left[ {7\over 3} \pi \alpha R - {4\pi R
\over \alpha}(\delta - {1\over 12}) \right]m^2 {\Gamma (s-1)
\over \Gamma (s-1/2)} + \left[ 2\pi^{3/2} (\delta - {1\over
12}) \right]m \nonumber \\
&+&\left. \left[{\pi\alpha \over R} (\delta - {1 \over
12} + {229 \alpha^2 \over 24\cdot 105}) - {2\pi \over \alpha R}
(\delta^2 - {\delta \over 6} + {7 \over 240}) \right]
{\Gamma (s) \over \Gamma (s- 1/2)} + \dots \right\}\ . \nonumber
\enn
Here $\delta = 2(1-\alpha^2)(\xi -1/8)$. These are all pole
contributions in zeta function, all next terms will be finite
for $s\to 0$. Comparing this expression with that obtained by
the Mellin transformation over trace of heat kernel (in three
dimensions) 
\bnn
\zeta_A(s-1/2) &=& {1 \over \Gamma (s-1/2)} \int_0^\infty dt
t^{s-3/2} K(t) = {m^{-2s} \over (4\pi )^{3/2}} \left\{ B_0 m^4
{\Gamma (s-2) \over \Gamma (s-1/2)} \right.\\
&+& \left. B_{1/2} m^3 {\Gamma (s-3/2)
\over \Gamma (s-1/2)} + B_1 m^2 {\Gamma (s-1) \over \Gamma
(s-1/2)} + B_{3/2} m + B_2 {\Gamma (s) \over \Gamma (s-1/2)} +
\dots \right\} \nonumber
\enn
we obtain the heat kernel coefficients : 
\bnn
B_0 &=& {4\pi R^3 \over 3\alpha}\ ,\ B_{1/2}= -2\pi^{3/2}R^2\ ,\
B_1 = {7\over 3} \pi \alpha R - {4\pi R \over \alpha}(\delta -
{1\over 12})\ ,\  \\ 
B_{3/2}&=& 2\pi^{3/2} (\delta - {1\over 12})
\ ,\ B_2 = {\pi\alpha \over R} (\delta - {1 \over
12} + {229 \alpha^2 \over 24\cdot 105}) - {2\pi \over \alpha R}
(\delta^2 - {\delta \over 6} + {7 \over 240}) \ .\nonumber
\enn
Those terms which are proportional to inverse degree of $\alpha$
come from exponential part of the uniform expansion
(\ref{UniformExpansion}), and respectively for $T(s)$ (\ref{T(s)}).
Terms which are linear on $\alpha^1$ or $\alpha^0$ come from the
series $\sum u_k/\mu^k$ in (\ref{UniformExpansion}). 

Now we may compare our results with well-known formulas given in
Refs. \cite{KennedyCritchleyDowker,ElizaldeBook,BransonGilkeyVass}. 
The coefficients $B_0,\ B_{1/2},\ B_{3/2}$ coincide with general
formulas in three dimensions (all geometrical quantities are given in
Sec.\ref{Geometry})
\bn
B_0 &=& {4\pi R^3 \over 3\alpha}= \int_V dV\ , \label{B01/23/2} \\
B_{1/2} &=& -2\pi^{3/2} R^2= -{\sqrt{\pi} \over 2} \int_{\partial
V} dS\ ,\ \nonumber \\
B_{3/2} &=& -4\pi^{3/2} (1 - \alpha^2)({1 \over 6} - \xi) -
{\pi^{3/2} \alpha^2 \over 6} \nonumber \\
&=& -{\pi^{1/2} \over 192} \int_{\partial V}\left( -96\xi{\cal
R} + 16 {\cal R} + 8 {\cal R}_{ik}N^iN^k + 7(trK)^2 - 10trK^2
\right) dS\ . \nonumber
\en 
We devide the coefficient $B_1$ into volume and boundary
contributions, according with general formulas and an additional
topological contribution $B_1^{top}$ which has not obtained
before 
\be
B_1 = \int_V b_1(x)dV + \int_{\partial V} c_1(x)dS + B_1^{top}\
, \label{B1}
\ee
where
\bd
b_1(x) = \left({1\over 6} - \xi\right){\cal R}\ ,\ c_1(x) = {1
\over 3}trK\ ,\ B_1^{top} ={\pi R \over 3}\left( {1\over \alpha
} - \alpha \right)\ .
\ed
Some problems are connected with the term $B_2$ because in this
case the volume part of $B_2$ \cite{Christensen} 
\bd
b_2(x) = -{1 \over 180} {\cal R}^{ik}{\cal R}_{ik} + {1 \over
180}{\cal R}^{iklj}{\cal R}_{iklj} + {1\over 6}\left({1\over 5}
- \xi\right) {\cal R}_{;k}^k + {1\over 2}\left({1 \over 6}
-\xi\right)^2{\cal R}^2 \ ,
\ed 
is proportional to $1/r^4$ and the integral over volume will
divergent at origin. This problem has been already discussed by
Cheeger \cite{Cheeger}, and Br\"uning and Seeley \cite{Bruning}
using {\em partie finite} of the integral. Then we have good
agreement with general formulas (for boundary contributions
$c_2$ see Ref. \cite{BransonGilkeyVass}) 
\bn
B_2 &=& \int_V Fb_2(x)dV + \int_{\partial V}Fc_2dS \label{B2}\\
&=&-{4\pi \over \alpha R} \left( 2(1-\alpha^2)^2 \left({1
\over 6}-\xi \right)^2 + {2\alpha^2 \over 3}(1 - \alpha^2)
\left( {1\over 5} - \xi\right) + {(1 - \alpha^2)^2 \over 90}
\right) - {16\pi \alpha^3 \over 315 R}. \nonumber 
\en 
\section{Ground State Energy}\label{GrEn}
Using the results of previous section and Eq.(\ref{GroundEnergy})
we obtain the following formula for regularized ground state
energy of massive scalar field in the global monopole background
\bnn
E(s) 
&=& {1\over 2} M^{2s}\zeta_A(s-1/2) \\
&=& \left({M\over m}\right)^{2s} {1\over 2(4\pi )^{3/2}}
\left\{ B_0 m^4 {\Gamma (s-2) \over \Gamma (s-1/2)} 
+ B_{1/2} m^3 {\Gamma (s-3/2) \over \Gamma (s-1/2)} \right.
\nonumber \\
&+& \left. B_1 m^2 {\Gamma (s-1) \over \Gamma
(s-1/2)} + B_{3/2} m + B_2 {\Gamma (s) \over \Gamma (s-1/2)} +
\dots \right\}\ , \nonumber
\enn
where the heat kernel coefficients $B_k$ are given by expressions 
(\ref{B01/23/2}), (\ref{B1}) and (\ref{B2}). 

This series has the form the Schwinger - DeWitt expansion with
additional topological contribution term in $B_1$ given by
Eq.(\ref{B1}). All divergencies of the energy for $s\to 0$ are
contained in these first five terms (three terms in the case without
boundary \cite{BD}). The next terms are finite in the limit $s\to 0$
and they have the following structure 
\bd
\sum_{k=1}^\infty {D_k \over m^k R^{k+1}}\ .
\ed
After renormalization we have to obtain ground state energy which
must obey the relation \cite{Method2} (see also
Refs. \cite{mtoinfty}) 
\bd
\lim_{m\to \infty} E^{ren} = 0\ .
\ed
For this reason for  renormalization we subtract first five terms 
and because we are interested in the quantum effects which don't
depend on the boundary we must turn $R$ to infinity. Obviously at the
end of the calculation, the ground state energy will be zero. 
\section{Discussion and Conclusion}\label{Last}
In this paper we have investigated the zeta function of the
Laplace operator and the ground state energy of massive scalar
field on the global monopole background. In the calculations we
have considered pointlike global monopole and took the metric in
the form given by Eq.(\ref{Metrica}). 

First of all we calculated zeta function of the Laplace operator
$\hat A = -\triangle + \xi {\cal R} + m^2$ on three dimensional
section ($t=const$) of global monopole spacetime with Dirichlet
boundary condition for the field on the surface of sphere of
radius $R$. We have rederived first five heat kernel coefficients
(\ref{B01/23/2}), (\ref{B1}) and (\ref{B2}) which were firstly
obtained in Ref. \cite{BordagKirstenDowker}. It is worth to
mention that the term $B_1$ contains an additional contribution,
$B_1^{top}$, which is due to the nontrivial topology of the
spacetime itself, besides the usual volume and boundary parts.
The topological contribution has the form below  
\bd
B_1^{top} ={\pi R \over 3}\left( {1\over \alpha } - \alpha
\right). 
\ed
As far as we know this is the first time where the purely
topological effects could be separated from the geometrical one,
in the calculation of the energy. As a matter of fact the
parameter $\alpha$ is related with the topology as well as with
the curvature.

The above expression is in agreement with the case of a conical
spacetime. Indeed, for $t=const$ and $\theta = \pi /2$,
Eq.(\ref{Metrica}) has the conical structure 
\bd
ds^2 = \alpha^{-2} dr^2 + r^2 d\varphi^2\ .
\ed
After trivial coordinate changing $r = \rho\alpha\ , \alpha =
1/\nu$ our topological term coincides with a similar one in conical
spacetime (see for example \cite{KhusnutdinovBordag}). We have 
to take into account only that in two dimensions coefficient
$B_1$ is dimensionless and we must drop the radius $R$ from
$B_1^{top}$. 

Next we obtained an expression for regularized ground state
energy of massive scalar field with arbitrary nonconformal
coupling $\xi$. In the zeta regularization approach the ground
state energy is proportional to the zeta function of the Laplace
operator on three dimensional spatial section of global monopole
manifold. Taking into account usual renormalization prescription
\cite{BD} we obtain that the ground state energy is zero.  This
result coincides with that in cosmic string spacetime
\cite{KhusnutdinovBordag}. It is easy to understand this from
dimensional consideration. In both cases there are no dimensional
parameters. There are only two dimensionless parameters $\nu = 1 -
4G\mu/c^2$ and $\alpha^2 = 1 - 8\pi G\eta^2/c^2$ for cosmic string
and global monopole spacetime, respectively. Parameters $\mu$ and
$\eta^2$ have dimensions of mass per unit lengh. Therefore, the
ground state energy has the following structure 
\bd
E = {\hbar c \over f(m,c)}\ ,
\ed
where the function $f$ with dimension $L$ depends on mass $m$
and velocity of light (gravitational constant $G$ has already
absorbed by dimensionless parameters $\nu$ and $\alpha$).
Obviously there is no way to construct quantity with
dimension of length using only $m$ and $c$. In the case of
energy density there is natural variable with dimension of
length, this is the radial coordinate $r$ and that is why one can
construct some expression for energy density. In the case of
global monopole with non zero core there will be a parameter with
dimension of length, namely size of core and we may expect that
ground state energy will depend on this parameter. But this
question should be investigated in a separate paper. 
\section*{Acknowledgments}
NK would like to thank Dr. M. Bordag for many helpful
discussions and comments, Dr. M. Volkov for a critical reading
of the paper, also NK is grateful to Departamento de
F\'{\i}sica, Universidade Federal da Paraiba (Brazil) where this
work was done, for hospitality. His work was supported in part
by CAPES and in part by the Russian Found for Basic Research,
grant No 97-02-16318.  

ERBM and VBB also would like to thank the Conselho Nacional de
Desenvolvimento Cientifico e Tecnol\'ogico (CNPq).

\end{document}